\documentclass[12pt,cites,graphicx]{article}

\usepackage{epsfig}

\title{A simple harmonic model as a caricature for mismatch and relaxation effects
for ion hopping dynamics in solid electrolytes}

\author{Magnus Kunow and Andreas Heuer\\[3mm]
Institut f\"ur Physikalische Chemie and Sonderforschungsbereich 458,\\[3mm] Westf\"alische Wilhelms-Universit\"at, Schlossplatz 4/7, 48149 M\"unster, Germany
}

\begin{document}

\maketitle

\renewcommand{\thefootnote}{\fnsymbol{footnote}}

\noindent

{\bf \large  Abstract}

We formulate a simple harmonic mean-field model with N+1 particles
and analyse the relaxation processes following a jump of one of
these particles. Either the particle can jump back
(single-particle route) or the other N particles adjust themselves
(multi-particle route). The dynamics of this model is solved
analytically in the linear response regime. Furthermore we relate
these results to a phenomenological approach by Funke and
coworkers (concept of mismatch and relaxation: CMR) which has been
successfully used to model conductivity spectra in the field of
ion dynamics in solid electrolytes. Since the mean-field model
contains the relevant ingredients of the CMR-approach, a
comparison of the resulting rate equations with the CMR-equations
becomes possible. Generalizations beyond the mean-field case are
discussed.

\section{Introduction}

\label{intro}

The complexity of ion dynamics in disordered systems is reflected
by the strong frequency dependence of the conductivity
$\sigma(\nu)$, observed at low temperatures \cite{pccpall}. The
dispersion can be directly interpreted as the presence of
correlated backward-forward dynamics of the mobile ions
\cite{Funkeenc,Roling01,Heuer02}. From a theoretical point of
view, two different mechanisms may contribute to the complex
dynamics. First, the ions move in an energy landscape, supplied by
the basically immobile network and which for amorphous
electrolytes is expected to be disordered. Second, the ions
interact via Coulomb interaction which tries to repel the ions
from each other. Furthermore, on longer time scales and at lower
temperatures the ionic dynamics can be described as hops between
adjacent ionic sites \cite{Kieffer94,Smith95,Kunow02}.

A theoretical description of ion dynamics on longer time scales
can therefore be restricted to hopping dynamics. So far, however,
the  problem of ion dynamics in solid electrolytes is too
complicated to formulate an analytic microscopic theory.
Neglecting, however, one of the two ingredients, significant
progress can be made. Without the interaction among the ions it is
possible to extract information about the ion dynamics via
percolation theory or effective medium theory\cite{ref1,Maass01}.
More specifically, in most cases a random energy landscape or a
random barrier landscape has been analysed \cite{Baranovskii99}.
 Not all predictions are compatible with
experimental data. Experimentally, it is observed that the length
scale $L$ on which the ion dynamics can be described as a random
walk is of the order of the nearest neighbor distance of two
typical ionic sites and is temperature independent
\cite{Roling01}. In contrast, for the random barrier model
 a significant temperature dependence as well as much
longer values of $L$ are observed \cite{Roling00}.

Other workers have stressed the relevance of Coulomb interaction
among the mobile ions. Analytical calculations \cite{ref9} and
simple model systems have been constructed along this line
\cite{ref10,Reinisch02}. In particular Funke and coworkers have
devised phenomenological models which are based on the relevance
of Coulomb interaction  \cite{ref2,ref3,ref4,ref5,ref8,Funke02}.
They consider the interaction among the ions as the relevant
ingredient, incorporating also effects of stress and strain. The
picture is as follows: after a jump of a well-equilibrated ion at
site A to an adjacent empty site B the new ionic environment will
often lead to an energetic mismatch, i.e. could be quite
unfavorable. Thus it is very likely that the ion jumps back to A
({\it single-particle route}). If, by chance, the backjump does
not occur immediately, the adjacent ions may have time to adjust
to the new situation and the ionic site B becomes more favorable.
Thus the backjump probability decreases with time. If finally site
B has become more favorable than the initial site A the jump can
be viewed as successful ({\it multi-particle route}). This
interplay between both routes gives rise to the complex ionic
dynamics, involving many correlated back- and forth jumps. Funke
has formulated two coupled differential equations, involving two
different functions \cite{ref8, Funke02}. One function basically
describes the mean square displacement of the individual ions, the
second function the response of the neighbor ions to an ionic
jump. In the previous version of the model (concept of mismatch
and relaxation: CMR) this coupled set has been solved analytically
and agrees well with measured conductivity spectra \cite{ref8}. In
the original CMR-approach no adjustable parameters are present
except for the simple scaling of the time and the length scale.
Presently, Funke and coworkers work on a more refined model with
one adjustable parameter to further improve the agreement with
experimental data \cite{Funke02}.

So far, the CMR-equations have not been derived from first
principles. Unfortunately, it is not possible to verify the
approach from comparison with experiments. The good agreement of
the mean square displacement with experiment (see below for a
closer discussion) is in a strict sense only a {\it necessary}
condition for the validity of the CMR-approach. Therefore, we feel
that a thorough theoretical discussion of the CMR-equations may be
valuable. The CMR-approach casts the interplay of single-particle
and multi-particle relaxation into two rate equations, using
phenomenological arguments. In particular, the precise nature of
the  interaction among the ions does not enter these equations.
The scope of the present work is to analyse a very simple but
non-trivial model which leads to exactly the effects, incorporated
into the CMR-approach. It will be shown that indeed two equations
can be derived with a formal structure identical to the
CMR-equations. A more detailed analysis, however, will reveal
differences of the exact solution as compared to the
CMR-equations. The effect of additional disorder will be also
taken into account. Furthermore it will be argued that the present
modifications of the CMR-equations can be justified also from a
strictly theoretical point of view.

The organization of this paper is as follows. In Section 2 we
introduce the CMR-approach. In Section 3 our model is introduced
which is then exactly solved in Section 4 in the limit of
vanishing disorder. Section 5 contains the derivation of the two
coupled rate equations which are formally identical to the
CMR-equations. Numerical simulations of the model, partly also
including disorder effects, are presented in Section 6. We end
with a discussion and a summary in Section 7.

\section{CMR: a short summary}

The CMR-approach is based on two central quantities $W(t)$ and
$g(t)$. $W(t)$ is formally defined as the normalized derivative of
the single-particle mean square displacement $(W(0) = 1)$. As
already mentioned above the key idea of the CMR-approach is to
consider the mismatch generated by a hop of a central ion (at time
t=0). In the multi-particle route the neighbors rearrange and thus
adjust to the new position of the central ion. In the
single-particle route the central ion jumps back.

The multi-particle relaxation is characterized by the function
$g(t)$. It expresses the normalised distance between the actual
position of the ion and the position at which it would be
optimally relaxed. During this relaxation process the central ion
is supposed to stay at its position so that it can be viewed as
fixed \cite{Funke02}. Due to the multi-particle relaxation the
actual position of the central ion gets closer to its optimum
position as determined by the neighbor positions so that the
energetic mismatch is slowly released. The initial situation after
the jump is characterized by $g(t=0)=1$. Finally $g(t)$ approaches
0 when the initial mismatch has decayed. The value -$\dot{g}(t)$
is thus proportional to the rate of mismatch relaxation along the
many-particle route.

Let $S(t)$ denote the fraction of ions which have not performed a
correlated backjump after time $t$, i.e. still correspond to the
successful jumpers. Then $-\dot{S}(t)/S(t)$ can be interpreted as
the rate of mismatch relaxation on the single-particle route.  The
factor $1/S(t)$ takes into account that a relaxation process at
time $t$ of a successful ion requires that the ion has remained at
the new position up to time $t$. The central assumption of the CMR
is that the single-particle and the multi-particle rates are
proportional to each other at all times, i.e. (CMR-Ia)
\begin{equation}
\label{eqcmr1a}
 - \dot{S}(t)/S(t) \propto  - \dot{g}(t).
\end{equation}

Furthermore Funke assumes that the single-particle rate can be
also expressed via (CMR-Ib)
\begin{equation}
\dot{S}(t)/S(t) \propto \dot{W}(t)/W(t).
\end{equation}
This means that the rate of mismatch relaxation on the
single-particle route is proportional to $-\dot{W}(t)/W(t)$. For a
random-walk with no correlated backward jumps one obviously has
$S(t) = 1$. Furthermore due to the strictly diffusive behavior one
also has $W(t) = 1$. Here CMR-Ib is trivially fulfilled. Combining
CMR-Ia and CMR-Ib one finally gets (CMR-I)
\begin{equation}
\label{eqwcmr1}
 - \dot{W}(t)/W(t) \propto  - \dot{g}(t).
\end{equation}

Next a rate equation for $g(t)$ is formulated. Here it is argued
that the decay of $g(t)$ is proportional to the convolution of the
driving force and the velocity autocorrelation function of the
neighboring mobile ions. Noting that the derivative of $W(t)$ is
proportional to the velocity autocorrelation function and assuming
that $W(t)$ decays much faster than $g(t)$ (this can be afterwards
 checked in a self-consistent manner) one ends up with (CMR-II)
\begin{equation}
\label{eqg} -\dot{g}(t) \propto  W(t) g(t).
\end{equation}
In the most recent version of the CMR Funke and coworkers have
included another time-dependent function which may represent the
time-dependent effective number of mobile neighbors available for
the relaxation \cite{Funke02}, i.e. (CMR-IInew)
\begin{equation} \label{eqgneu}
-\dot{g}(t) \propto  W(t) g(t)n(t).
\end{equation}
 From the fitting it turns out that the choice $n(t) =
g(t)$ yields a very good agreement with the experimental data. In
the latter part of this manuscript we briefly discuss this
modification of CMR-II.

These relations  form a closed system. Note that all relations are
based on general arguments. If CMR-I and CMR-II were valid they
should be applicable to a large class of models for which the
relaxation occurs by the interplay of single-particle and
multi-particle routes. The function $W_{CMR}(t)$, obtained from
the solution of CMR-I and CMR-II, can finally be compared with
experimental conductivity spectra and thus with the experimental
$W(t)$.

Now we are in a position to discuss why the empirical agreement of
$W_{CMR}(t)$ with the experimental $W(t)$ does {\it not} imply
that the CMR-equations are correct. Validation of the
CMR-equations naturally requires knowledge of $g(t)$. The
definition of $g(t)$, however, is based on the specific situation
of a particle fixed after its jump. Whereas $g(t)$ can be
determined from computer simulations or in some cases even
analytically (see below), no experimental access is presently
available. In principle, it may turn out that both CMR-I and
CMR-II are not correct but the final solution of $W_{CMR}(t)$ can
be used to describe experiments. In any event, the goal of this
work is to analyse CMR-Ia, CMR-Ib,  CMR-II and later on also
CMR-IInew individually.

\section{Harmonic model}

We want to construct a possibly very simple model which allows for
the presence of a single-particle and a multi-particle route. The
single-particle route requries an interaction term which is a
minimum when adjacent particles have their respective equilibrium
distance. This can be most easily achieved by a harmonic
potential. The multi-particle route is facilitated by a
translationally invariant model. After the jump of a particle and
the consecutive jumps of all adjacent particles in the same
direction the original configuration is recovered. For reasons of
simplicity one may take a mean-field model in which all particles
are identical.

Both properties are contained in  the Hamiltonian
\begin{equation}
H = (k/N) \sum_{i=1}^N \sum_{j > i}^N (y_i - y_j)^2 + (k/N)
\sum_{i=1}^N (y_i - x)^2 \equiv H_y + H_{xy}.
\end{equation}
 The ground state of this system is $x=y_1=...=y_N$. Of course,
at finite temperature there will be a spread of the $y_i$ which
can be easily quantified by minimization of the free energy (not
shown in this paper). All $(N+1)$ particles are treated
identically in this mean-field Hamiltonian. The particle with
coordinate $x$ is formally viewed as the central particle and we
define it as the {\it x-particle}. Furthermore we define the
center-of-mass $y$ of the other $N$ particles as
\begin{equation}
y = (1/N) \sum_{i=1}^N y_i.
\end{equation}
The set of particles $y_1,...,y_N$ will be denoted {\it y-cloud}.
If after the jump $x \ne y$,  the x-particle will experience a
back-dragging force until $x= y$. On the one hand, this can be
achieved by a jump of the x-particle to this favorable position
(single-particle route). On the other hand, the y-cloud can adjust
to the new position of the x-particle (multi-particle route).

In order to mimic the hopping dynamics the coordinates $y_i$ of
the individual particles are discretised ($y_i =
...,-3/2,-1/2,1/2,3/2,...$). Without interaction among the
particles ($k=0$) the rates for all jump processes are identical.
With finite interaction the rates have to reflect the energy
variation due to the transition. In the spirit of the Metropolis
criterion the rate for a jump process of one particle is
unmodified as compared to the $k=0$ case if the energy decreases
due to the jump and is decreased by $\exp(-\beta \Delta U)$ if the
energy increases by $\Delta U$ \cite{ref15,ref16}.

In this model it is possible to take into account disorder
effects, as present in amorphous ion conductors for which the
CMR-approach has been applied as well. Here we restrict ourselves
to barrier disorder. For the barriers between any two sites
$(a,a+1)$ of the i-th particle we choose a random but fixed value
$E_i(a,a+1)$, equally distributed between 0 and $V_{max}$. Then
the individual transitions are modified by multiplying every rate
with the factor $\exp(-\beta E_i(a,a+1))$ if a transition of the
i-th particle from $y_i=a$ to $y_i = a + 1$ (or vice versa) is
considered. The discretised positions as well as the barrier
disorder are sketched in Fig.1. For $k
> 0$ the particles are confined to a finite region of $y_i$ values,
as also indicated  in Fig.1.

For comparison with the two CMR rate equations we need to express
$W(t)$ and $g(t)$ in terms of the system coordinates. Formally, we
define
\begin{equation}
W(t) = w(t)/w(0)
\end{equation}
where $w(t)$ is the derivative of the single-particle mean square
displacement
\begin{equation}
w(t) \equiv (d/dt) \langle (x(t) - x(0))^2 \rangle.
\end{equation}

For $k=0$ and without disorder one would have a simple random walk
of the x-particle, i.e. $\langle (x(t)-x(0))^2\rangle = \Gamma t$
and thus $w(t) = \Gamma$ where $\Gamma$ is a rate constant.  With
additional disorder ($V_{max} \ne 0$) the short-range dynamics is
still diffusive albeit with a smaller rate $\Gamma$, reflecting
the slowing-down due to the presence of higher barriers.

It is possible to write $W(t)$ in a more elegant way which will be
of relevance later on. Let $\langle . \rangle_0$ denote the
ensemble average over all events where at time $t=0$ the
x-particle jumps one site to the {\it right}. We start from the
general relation (valid for $t > 0)$
\begin{equation}
2 \langle \dot{x}(0)\dot{x}(t) \rangle = \dot{w}(t) = \Gamma
\dot{W}(t).
\end{equation}
The term in the brackets of the l.h.s. is non-zero if there is a
jump happening at $t
= 0$. The probability for such a jump during a time interval
$\Delta t$ is $\Gamma \Delta t$. Using a discrete notation $
\dot{x}(0)= (x(\Delta t/2) - x(-\Delta t/2))/\Delta t$ one can
write in case of a jump: $ \dot{x}(0) = 1/\Delta t$. This yields
\begin{equation}
\langle \dot{x}(0)\dot{x}(t) \rangle = (\Gamma\Delta t) \langle
(1/\Delta t)\dot{x}(t) \rangle_0 = \Gamma\langle \dot{x}(t)
\rangle_0
\end{equation}
and thus
\begin{equation}
\label{eqwdot}
\dot{W}(t) = 2 \langle \dot{x}(t) \rangle_0.
\end{equation}
In order to obtain $W(t)$ from this relation one has to recognize
that for reasons of time reversal symmetry
\begin{equation}
\langle x(0_+) - y(0)\rangle_{after \, jump}
= - \langle x(0_-) - y(0)\rangle_{before \,  jump}
\end{equation}
where $x(0_\pm)$ denotes the position of the x-particle after and
before the jump at $t=0$, respectively.  This relation expresses
the fact that before a jump to the right the x-particle is
typically left to the center of mass of the y-cloud and after a
jump by the same amount right to it. Since the jump length is unity
one can directly conclude
\begin{equation}
\label{eqxy}
\langle x(0_+) - y(0)\rangle_0 = 1/2.
\end{equation}
Integration of Eq.\ref{eqwdot} under the condition $W(0) = 1$ thus
yields
\begin{equation}
\label{eqw}
W(t) = 2 \langle x(t) - y(0) \rangle_0.
\end{equation}
On a qualitative level this relation shows that the mean square
displacement can be obtained from analysing the response of the
system after a jump at $t=0$.

For later purposes we also consider the multi-particle mean square
displacement $\langle (y(t) - y(0))^2\rangle$ if the x particle
were not present, i.e. exclusively for the Hamiltonian $H_y$. In
analogy to $W(t)$ we define $w_y(t)$ as its derivative and
$W_y(t)$ as its normalized derivative. Since for short times the N
particles behave independently of each other the short-time
diffusion constant of the center of mass is scaled by a factor of
$1/N$ as compared to the single-particle case. Thus $w_y(0) =
w_x(0)/N = \Gamma/N$. $w_y(t)$ can be simply calculated for the
case of no disorder. Since all interaction terms  cancel out, the
center of mass of the y-cloud will just perform a random walk,
i.e. $w_y(t) = \Gamma/N$. In contrast, in the presence of disorder
the dynamics of the center of mass can be highly non-diffusive.
Later on it will be important that exactly for the case of
diffusive dynamics the sum over all displacements during one time
step is independent of the sum over all displacements at a
previous time. In analogy we introduce $w_x(t)$ as the derivative
of the mean-square displacement of the x-particle if no
interaction with the y-cloud is present.

As a next step, we define $g(t)$ as the relaxation of the y-cloud
for
 a fixed x-particle with the normalization condition $g(0) = 1$. This definition is
 motivated by the use of $g(t)$ in the CMR-approach.
 Let $\langle . \rangle_F$ denote the ensemble
 for which at time t=0 the x-particle has jumped to the right and remains
 there {\it fixed}
 forever. Since on average $\langle x - y(0) \rangle_F = 1/2$
 the appropriate definition is
 \begin{equation}
 \label{eqg-1}
 1-g(t) = 2\langle y(t) - y(0) \rangle_F.
 \end{equation}
 $g(t)$ describes how the y-cloud approaches the fixed position of
 the x-particle, corresponding to a decay of $g(t)$ from 1 to 0.
 Note that an equivalent expression is given by
\begin{equation}
\label{gsimple}
g(t) =  2 \langle x - y(t) \rangle_F.
\end{equation}

\section{Analytic solution of the model}

In this section we calculate the particle dynamics in this model
without disorder. We start with the analysis of the dynamics of the
y-cloud (first, without the x-particle, i.e. for the Hamiltonian
$H_y$) under some perturbation. In case of a time-dependent force
$F(t)$, i.e. an interaction Hamiltonian $F(t) y$, one can apply
linear response theory for small $F(t)$ to express the dynamics of
$y$ under the perturbation in terms of equilibrium correlation
functions, yielding \cite{ref17}
\begin{equation}
\label{eqlinres}  \langle \dot{y}(t) \rangle_{pert} = \beta
\int_0^t d\tau F(\tau) \langle \dot{y}(t) \dot{y}(\tau) \rangle
\approx (1/2)\beta F(t) w_y(t)
\end{equation}
with $\beta = 1/(k_B T)$. The final approximation is valid if the
time-dependence of the force is much slower than the decay of the
velocity-velocity autocorrelation function.

In the present case we have no fixed force $F$ but the perturbation
energy for a fixed x-particle is given by $H_{xy}$. It can be
rewritten as
\begin{equation}
H_{xy} = k (y - x)^2 + (k/N) \sum_{i=1}^N (y - y_i)^2
\end{equation}
The second term  is independent of $x$. Its average value results
from a free energy minimization of the positions of the particles
which are part of the y-cloud. Thus it is irrelevant for the
perturbation. For given $y$ and $x$ the force $F$ thus reads $2k
(x - y)$ which shifts the y-cloud towards the x-particle. For not
too small $N$ it is evident from Eq.\ref{eqlinres} that the
response of the y-cloud is slow due to the factor $1/N$, which is
contained in $w_y(t)$. Thus one can use the approximation in
Eq.\ref{eqlinres} and finally obtains (omitting the index {\it
pert})
\begin{equation}
\label{eqydot} \langle \dot{y}(t) \rangle = k\beta (x -
y(t))w_y(t).
\end{equation}
Furthermore, the linear response relation Eq.\ref{eqydot} is also
valid if the x-particle is allowed to be mobile, i.e. $x$ is
substituted by $x(t)$. Thus one has, using the $\langle .
\rangle_0$-ensemble,
\begin{equation}
\label{eqrate1} \langle \dot{y}(t) \rangle_0 = \beta k w_y(t)
(\langle x(t) \rangle_0
 - \langle y(t) \rangle_0).
\end{equation}
Note that this relation also holds for disorder, since the
disorder is taken into account by the term $w_y(t)$. Without
disorder one may insert $w_y(t) = \Gamma/N$.

In analogy one can also formulate how the x-particle is attracted
by the y-cloud. For this purpose we consider an isolated
x-particle which is distorted by the energy $F(t) x$ with $F(t) =
-2k (x - y)$. We obtain
\begin{equation}
\label{eqrate2} \langle \dot{x}(t) \rangle_0 = \langle \dot{x}(t)
\rangle_{0,do} + (\beta k w_x(t)) (\langle y(t) \rangle_0
 - \langle x(t) \rangle_0).
\end{equation}
The function $w_x(t)$ characterizes the dynamics of an isolated
x-particle. The first term on the right side corresponds to the
behavior of the x-particle if no coupling to the y-cloud were
present. Without disorder one has $w_x(t) = \Gamma$ (random-walk)
and $\langle \dot{x}(t) \rangle_{0,do} = 0$. With disorder
$w_x(t)$ may display a complex time-dependence.

This set of rate equations can be solved in a straightforward way.
With the initial conditions $\langle x(0) \rangle_0 = 1/2$ and
$\langle y(0) \rangle_0 = 0$ one obtains (using the abbreviation
$C = \beta k $)
\begin{equation}
W(t)/2 = \langle x(t) \rangle_0 = \frac{1}{2(1+N)} \left ( 1 + N
\exp(-C \Gamma t) \right )
\end{equation}
and
\begin{equation}
\langle y(t) \rangle_0 = \frac{1}{2(1+N)} \left ( 1 - \exp(-C
\Gamma t) \right ).
\end{equation}
Furthermore one can easily calculate $g(t)$ from Eq.\ref{eqrate1}
by keeping the x-particle fixed. Together with Eq.\ref{gsimple}
one obtains
\begin{equation}
\label{eqsolg}
 g(t)= \exp(-C \Gamma t/ N).
\end{equation}
In agreement with typical experimental situations the decay of
$g(t)$ is much slower than the decay of $W(t)$.

Interestingly, on the level of the linear response  the fact that
the y-cloud is composed of individual particles is lost. Rather
one could have started from the very beginning with a two particle
problem, i.e. with the Hamiltonian $H = k(x - y)^2$ and postulate
that for $k=0$ the equilibrium dynamics of the x-particle is
characterized by $w_x(t)$ and that of the y-particle by $w_y(t)$.
This simplified view will be of importance later on.

For checking the validity of the CMR-equations for the present
case $V_{max} = 0$ (no disorder) we will proceed in two steps.
First, we will {\it derive} rate equations in terms of $g(t)$ and
$W(t)$ for this Hamiltonian, thereby generalising the above
analysis to the case of additional disorder. Due to the strict
derivation all proportionality constants can be expressed in terms
of system parameters.  Second, comparison with CMR-Ia, CMR-Ib, and
CMR-II will reveal under which conditions the phenomenological
CMR-equations can be indeed applied. Third, we will check the
validity of our rate equations by comparison with numerical
simulations.

\section{Analytical derivation of CMR-like equations}

First we consider CMR-Ia. As mentioned before, without disorder
the center of mass of all particles performs diffusive dynamics.
This can be expressed more formally. Let $\Delta x(t)$ denote the
distance moved by the x-particle at time step $t$ ($\Delta x(t)$
can be either $-1,0,1$). In analogy, we define $\Delta y(t)$ as
the motion of the center of mass of the y-cloud at this time step.
Since the motion to the left and the right side are equally likely
one (trivially) has
\begin{equation}
\langle \Delta x(t)\rangle + \sum_{i=1}^N \langle \Delta
y_i(t)\rangle = \langle \Delta x(t)\rangle + N \langle \Delta
y(t)\rangle = 0
\end{equation}
 where the brackets indicate the ensemble average.

It turns out to be helpful to introduce the notation $\langle .
\rangle_{S}$. It denotes the average over the ensemble where at
time 0 the x-particle has jumped to the right and for times less
than time t has {\it on average} stayed at its new position.
Qualitatively, this is the ensemble of events for which the
initial jump at $t=0$ is  {\it successful} at least until the jump
at time t. Formally this means that $\langle x(t) \rangle_S =
\langle x(0_+) \rangle_0$. Because of the independence of the
dynamics during successive time steps for the purely diffusive
case (no disorder; see above) one also has
\begin{equation}
\langle \Delta x(t)\rangle_S + N \langle \Delta y(t)\rangle_S = 0.
\end{equation}

Obviously, the total system relaxes towards the equilibrium
situation in a non-oscillatory manner. Qualitatively, this implies
that those particles which are still successful at time $t$ (after
a jump to the right at $t=0$) will, on average,  have a tendency
to move to the left afterwards. Thus it is possible to iteratively
define the fraction $S(t)$ of successful particles, and thus the
S-ensemble, by two conditions: (i) the average value of $\Delta
x(t)\equiv x(t_+) - x(t_-)$ is zero for those particles which
remain in the S-ensemble after the jump at time t. (ii) the
average value of $\Delta x(t)$ is -1 for those particles which
were part of the successful ensemble before the jump at t and fall
out of the S-ensemble after the jump. Of course, this does not
imply that all particles jumping to the left, leave the S-ensemble
but only those which are not balanced by particles jumping to the
right. Therefore for a random walk, leading to purely diffusive
dynamics, one has $S(t) = 1$ because on average the fraction of
particles moving to the left and to the right is identical.  The
construction of the fraction of particles, belonging to the
S-subensemble, is sketched in Fig.2.

The definition of the S-subensemble implies
\begin{equation}
\langle \Delta x(t) \rangle_S = \frac{S(t-\Delta t) -
S(t)}{S(t-\Delta t)}\cdot (-1) + \frac{S(t)}{S(t-\Delta t)}\cdot 0
=  \frac{S(t) - S(t-\Delta t)}{S(t-\Delta t)}.
\end{equation}

In the limit $\Delta t \rightarrow 0$ one thus obtains
\begin{equation}
\frac{\dot{S}(t)}{S(t)} + N \langle \dot{y}(t) \rangle_S = 0
\end{equation}

For more general energy landscapes, e.g., with random barriers, the
center of mass dynamics is not simply diffusive. Rather a jump will
be typically followed by correlated backward dynamics. Thus the
presence of a jump of the x-particle at time $t=0$ to the right
implies that $\Delta x(0) + N \Delta y(0)$ on average is positive
and correspondingly will be negative at later times. Thus in
general one expects
\begin{equation}
\label{eqfdo}
 \frac{\dot{S}(t)}{S(t)} + N \langle \dot{y}(t)
\rangle_S = f_{do}(t) \le 0
\end{equation}
where $f_{do}(t)$ (do: disorder) represents the effect of disorder.

We can proceed further by using Eq.\ref{eqydot}. This relation has
important implications for our analysis. Due to the linearity of
the r.h.s. in x the average time-dependence of $y(t)$ is the same
whether one considers an ensemble where all x-particles are fixed
at some position $x_0$ or whether the x-particles are distributed
around this position with exactly the average value $x_0$. This
implies that the S-ensemble and the F-ensemble yield the same
time-dependence for the relaxation of the y-cloud. Thus
Eq.\ref{eqfdo} can be rewritten (using the relation $\dot{g}(t) =
- 2 \langle \dot{y}(t) \rangle_F = - 2 \langle \dot{y}(t)
\rangle_S $ from Eq.\ref{eqg-1}) as
\begin{equation}
\label{eqcmr1aneu}
 -\frac{(d/dt)S(t)}{S(t)} = -\frac{N}{2}\dot{g}(t) - f_{do}(t).
\end{equation}
Except for the disorder-term $f_{do}(t)$, which for weak disorder
may be small, relation CMR-Ia has been recovered.

Eq. \ref{eqydot} can be also used to derive CMR-II. After
expressing both sides in the F-ensemble and using Eq.\ref{gsimple}
as well as $\dot{g}(t) = -2\langle \dot{y}(t) \rangle_F $ one gets
\begin{equation}
\label{eqcmr2} \dot{g}(t) = k\beta g(t) w_y(t).
\end{equation}
The only difference to CMR-II is the substitution of the
single-particle quantity $w_y(t)$ by the multi-particle quantity
$w(t)$ (thus yielding the proportionality to $W(t)$). Since,
however, in typical experimental situations $w(t)$ and $w_y(t)$
are quite similar this substitution may be justified for practical
purposes.

It remains to check CMR-Ib. For this purpose we consider the
situation that at $t=0$ the x-particle jumps to the right side and
ends up at a position which, on average, is given by $y(0) + 1/2$
(see above). For reasons of simplicity we choose $y(0) = 0$. After
the next jump processes at $t = \Delta t$ there will be a higher
probability for the x-particle to jump to the left than  to the
right because of the back-dragging effect of the y-cloud. $p_1$
denotes the fraction of x-particles which effectively jump to the
left side (which means, which are not balanced by particles moving
to the right side; see above). Thus the number of successful
particles at $t = \Delta t$ is given by $S(\Delta t) = 1 - p_1$.
Furthermore this implies
\begin{equation}
\langle x (\Delta t) \rangle_0 = (1/2) (1 - p_1) + (-1/2) p_1.
\end{equation}
Together with Eq.\ref{eqw} one gets
\begin{equation}
W(\Delta t) = 1 - 2 p_1.
\end{equation}
Thus one has
\begin{equation}
\frac{W(0) - W(\Delta t)}{W(0)} = 2 \frac{S(0) - S(\Delta t)}{S(0)}.
\end{equation}
This relation implies that if CMR-Ib were valid one should choose
\begin{equation}
\label{eq1bexpl}
 \frac{\dot{W}(t)}{W(t)} =
2\frac{\dot{S}(t)}{S(t)}.
\end{equation}

 Now we analyse the next jump
process at $t = 2 \Delta t$. The x-particles, which are still
successful after $t = \Delta t$ and thus are still centered around
$x = 1/2$ will again have a tendency for a jump to the left side.
In analogy to $p_1$, we define $p_2$ as the fraction of these
x-particles, which effectively jump to the left side. In contrast,
the x-particles which were unsuccessful after the first jump and
are thus centered around $x = -1/2$ will also be attracted by the
y-cloud. For these particles this will result in a preference of
jumps to the right side.   In analogy to $p_2$ we define $q_2$ as
the fraction of these x-particles which effectively jump back to
$x = 1/2$. With these parameters one directly gets
\begin{equation}
S(2 \Delta t) = S(\Delta t) (1 - p_2)
\end{equation}
and
\begin{equation}
(1/2) W(2\Delta t) = \langle x (2 \Delta t) \rangle_0 = (1/2)
\cdot [(1 - p_1) (1 - p_2) + p_1 q_2] + (-1/2) [ p_1 (1 - q_2) +
(1-p_1) p_2].
\end{equation}
Straightforward algebra yields
\begin{equation}
\frac{W(\Delta t) - W(2\Delta t)}{W(\Delta t)} = 2 \cdot \frac{1 -
(1 + q_2/p_2) p_1}{1 - 2p_1} \cdot\frac{S(\Delta t) - S(2\Delta
t)}{S(\Delta t)} .
\end{equation}
Thus the proportionality of the normalised W- and S-derivative is
equivalent to the relation $q_2/p_2 = 1$. This relation is,
however, strongly violated at longer time-scales. The physical
origin of this violation is straightforward . Due to the
attraction of the x-particles the y-cloud will slowly shift to the
right, i.e. one expects $1/2 > y(\Delta t) > 0$. This implies that
x-particles at $-1/2$ will have a {\it stronger} tendency to jump
to the right than x-particles at $1/2$ to jump to the left.
Formally this can be expressed as $q_2 > p_1 > p_2$. Thus on times
scales for which the y-cloud starts to move to the right the
proportionality between $\dot{W}(t) /W(t)$ and $\dot{S}(t) /S(t)$
breaks down. Qualitatively, this means that $W(t)$ decays more
slowly than expected because those particles which have performed
a backward jump (after the initial jump) have a very strong
tendency to perform afterwards a forward jump. Note that for this
general analysis no relation to specific properties of the present
model was necessary.

We can explicitly check that CMR-Ib is indeed violated for longer
times when taking the analytical solution. If CMR-Ia and CMR-Ib
and thus CMR-I were valid the ratio $\dot{W}(t)/ ( W(t)
\dot{g}(t))$ should be constant for all times. Here we get
\begin{equation}
\label{eqexplicit} \frac{(N+1)\dot{W}(t)}{N^2 W(t) \dot{g}(t)} =
\frac{(N+1)\exp(-C\Gamma t(1-1/N)}{1 + N \exp(-C\Gamma t))} \equiv
D(t).
\end{equation}
Whereas $D(t)$ is $1$ for short times it approaches 0 for long
times. Thus in agreement with our general arguments $W(t)$ decays
slower than expected by CMR-Ia and CMR-Ib. More insight is gained
by rewriting $D(t)$ after a Taylor-expansion of $\ln D(t)$ in 1/N
as
\begin{equation}
\ln D(t) = \frac{1}{N}(1 + C\Gamma t - \exp(C\Gamma t)) \approx
-\frac{C^2\Gamma^2 t^2}{2N}
\end{equation}
This approximation is valid for $\exp(C\Gamma t)/N \ll 1$. Thus
the time-dependence of this term becomes relevant for
$C^2\Gamma^2t^2/N = O(1)$. On this time-scale $\tau$ one has
$W(\tau) = (1 + N \exp(-\sqrt{N})/(1+N)$ which for large $N$ is
already close to $W(t \rightarrow \infty)$. In agreement with our
general discussion the deviations occur if $g(t)$ starts to
decrease. At the crossover time $\tau$ one has $g(\tau) \approx 1
- 1/\sqrt{N}$.

 For a direct
visualisation of this effect we calculate $W_{CMR}(t)$. From CMR-Ia
and CMR-Ib together with Eq.\ref{eqsolg} one obtains after
appropriate normalization
\begin{equation}
W_{CMR}(t) = \exp[N(\exp(-C\Gamma t/N)-1)]
\end{equation}
$W(t)$ and $W_{CMR}(t)$ are compared in Fig.3. Since for $t <
\tau$ the time-dependence of $W_{CMR}(t)$ is close to the true
solution (for large $N$) the function $W_{CMR}(t)$ reproduces the
time-dependence of $W(t)$ up to the final plateau. This can be
explicitly seen in Fig.3.  Note that the plateau value of
$W_{CMR}(t)$ is $\exp(-N)$ which is much smaller than the limiting
value of $W(t \rightarrow \infty)=1/(1+N)$. The weak deviations of
$W(t)$ and $W_{CMR}(t)$ at short times result from a term
proportional to $1/N^2$ which has been neglected in the Taylor
expansion.

\section{Comparison with numerical simulations}

In this section we perform a detailed comparison of the different
rate equations of the CMR-appraoch, involving the functions $W(t)$
and $g(t)$,  with the outcome for our model system. From the
previous discussion we already know that CMR-Ia and CMR-II hold
without disorder whereas CMR-Ib is violated for long times. From
our general discussion we anticipate that in case of additional
disorder also CMR-Ia should be violated. Since no exact solution
is available with disorder we have performed kinetic Monte-Carlo
simulations of the hopping dynamics using the Metropolis
criterion. To check our analytical solution we have performed
simulations without and with disorder. We have chosen $N=64$ and
$C=1.23$ for most simulations.

The functions $w(t)$ and $w_y(t)$ can be easily extracted from the
simulated dynamics. For the calculation of $g(t)$ we used a
slightly modified simulation strategy. If during the simulation
the (randomly selected) x-particle has jumped to the right this
particle was excluded from further jumps for a fixed time
interval. During this time interval the relaxation of the other
particles were taken for the determination of $g(t)$. After this
time interval  this particle is again allowed to perform hopping
processes and another particle is selected as the x-particle and
so on. Averaging over a sufficient number of iterations one
obtains $g(t)$. During these runs we also calculated $\langle
\Delta x(t) \rangle_F$ and $\langle \Delta y(t) \rangle_F$. For
$\langle \Delta y(t) \rangle_F$ we simply determined the dynamics
of the y-cloud at time $t$ after the jump of the selected
x-particle. For the determination of $\langle \Delta x(t)
\rangle_F$ we calculated the probability that the x-particle {\it
would} jump at time $t$ either to the right or to the left. This
directly yields $\langle \Delta x(t) \rangle_F$, i.e. the average
variation of the position of the x-particle at time $t$ under the
condition that it was fixed after its initial jump at $t=0$. It is
evident that the determination of $g(t)$ as a multi-particle
quantity is more time-consuming than that of $W(t)$ as a
single-particle quantity.

For a direct comparison of the CMR-equations with simulated data
it turns out to be helpful to integrate the CMR-equations. CMR-I
and CMR-II yield
\begin{equation}
\label{eqcomp1}
 -\ln W(t) = \lambda_I (1 - g(t))
\end{equation}
and
\begin{equation}
\label{eqcomp2} -\ln g(t) = \lambda_{II} \langle r_y^2(t) \rangle,
\end{equation}
respectively, with  proportionality constants $\lambda_i$.
Comparison of CMR-II with Eq.\ref{eqcmr2} indicates that $\langle
r_y^2(t) \rangle$ would be a more appropriate choice for a
quantitative comparison. In any event, due to the similarity of
$\langle r_y^2(t) \rangle$ and $\langle r^2(t) \rangle$ for
typical experiments on ion conductors this modification does not
hamper the applicability of the CMR-approach
\cite{Haven65,Isard99}.

In Fig.4 we analyse the validity of CMR-I for the harmonic model
without disorder. The numerically determined functions $W(t)$ and
$g(t)$ agree very well with the respective analytical predictions.
As already discussed in Section 4 one expects deviations for long
times. For shorter times, however, $\ln W(t)$ and $1 - g(t)$ are
proportional to each other. Combination of Eq.\ref{eqfdo} and
Eq.\ref{eq1bexpl} shows that $\lambda_1 = N$ in agreement with the
simulated data .

The test of CMR-II can be seen in Fig.5. Here a perfect agreement
can be found for the full time regime. This was expected from
Eq.\ref{eqcmr2}. The predicted proportionality constant $\lambda_2
= C$ is also recovered.

Simulations with disorder ($V_{max} = 2.0$) can be found in Figs.6
and 7. One can see that CMR-I again displays significant
deviations (Fig.6).  CMR-II is fulfilled very well (Fig.7). Here
the proportionality constant $\lambda_{II}$ is close to the value
of $C$. The agreement for CMR-II does not come as a surprise since
we were able (see above) to derive CMR-II via linear response
theory.

Further insight about the effect of disorder can be obtained from
analysis of Eq.\ref{eqfdo} using again the original parameters
$C=1.23$ and $V_{max} = 2.0$. With disorder the center of mass of
the x-particle together with the y-cloud displays non-diffusive
dynamics, i.e. $f_{do}(t) < 0 $. This is explicitly shown in Fig.8
where we display $\langle \Delta x(t) \rangle_F$, $N \langle
\Delta y(t) \rangle_F$ and $f_{do}(t)$.  As anticipated, the
function $f_{do}(t)$ is negative. Already for $t = 10$ the
function $N\langle \Delta y(t)\rangle_F$ and thus
$-(N/2)\dot{g}(t)$ is much smaller than $-f_{do}(t)$. With
Eq.\ref{eqcmr1a} this implies that already for $t > 10$ the
dispersion of the x-particle as characterized by $\langle \Delta
x(t) \rangle_F$ is mainly determined by disorder rather than the
mismatch effect due to the y-cloud.

\section{Discussion and Summary}

We have presented a simple harmonic mean-field model which
contains single-particle as well as multi-particle relaxation
modes. The main goal was to check whether the CMR equations which
are based on the interplay between these two relaxation modes can
be derived for this model. First, CMR-II could be derived in
linear response theory after substituting $w_y(t)$ by $w(t)$.
Given the experimental similarity of $w_y(t)$ and $w(t)$ this step
may be justified. Second, CMR-Ia is valid without disorder. In
contrast, with disorder an additional term enters CMR-Ia which
accounts for the dispersive behavior related to disorder. Third,
CMR-Ib is only valid for short times. The physical reason for the
discrepancy at long times could be identified as the relaxation of
the y-cloud after the initial jump of the x-particle. Indeed, this
problem is inevitable if one wants to formulate a theory involving
a quantity like $g(t)$ where a particle is kept fixed and one
deals with the time-dependence of the fraction of successful
particles. In general it is  not possible to relate the behavior
of the successful particles to the mean square displacement since
the latter quantity also involves the behavior of those particles
which became unsuccessful at earlier times and may show a complex
time-dependence afterwards. Formally, this means that there is no
strict way to relate the $\langle . \rangle_0$-ensemble (where
$W(t)$ is defined) to the $\langle . \rangle_S$-ensemble (where
$S(t)$ and $g(t)$ are defined). Due to the generality of our
arguments the same problem would hold if one considers dynamics in
three rather than one dimension.

Maybe the most dramatic simplification of the model is the fact
that one particle interacts with all other particles in an
identical way. In reality the interaction strength depends (on
average) on the distance between two particles. To get a first
impression of this distance dependence one may introduce a step
function for the interaction strength in our model such that a
particle is only interacting with a fraction of the other
particles.  In the extreme limit this would coincide with the
Rouse model of polymer physics where each monomer only interacts
with the two adjacent monomers \cite{ref13,ref14}. For such a
model one may define the y-cloud such that it is composed of those
particles which directly interact with the (again randomly chosen)
x-particle (in the case of the polymer the two nearest neighbors)
and the z-cloud by the particles which do not interact with the
x-particles. Here we use the simplified picture for which we
forget that the cloud is composed of individual particles. As
shown in Section 4, this simplification was possible for the
mean-field case. Since we are only interested in a qualitative
discussion of this model extension we apply this simplification
also to the present non-mean field case.  Then our problem boils
down to a 3-particle problem with the Hamiltonian
\begin{equation}
H = k_{xy} (x - y)^2 + k_{yz}(y -z)^2.
\end{equation}
 Qualitatively, adding the effect of far away particles as expressed
 by the presence of the z-cloud, one expects
that at short times the decay of $g(t)$ is unmodified because the
only driving force comes from the shifted x-particle. In contrast,
at longer times the decay becomes slower because the z-cloud tries
to keep back the y-cloud. Somewhat related arguments have been
used to rationalize the additional $h(t)$-term in CMR-IInew as
compared to CMR-II \cite{Funke02}. For the above Hamiltonian we
can directly formulate the rate equations in analogy to Section 5.
To be more specific we consider the rate equations for this
Hamiltonian (without disorder), evaluated in the $\langle .
\rangle_S$-ensemble, to determine the effect of a fixed (or,
analogously, successful) x-particle
\begin{eqnarray}
\label{eqneu}
 \langle \dot{y}(t) \rangle_S &=& \beta w_y(t) [k_{xy} (
x - \langle y(t)
\rangle_S) + k_{yz}(\langle z(t) \rangle_S - \langle y(t) \rangle_S)] \\
\label{eqneu2}
 \langle \dot{z}(t) \rangle_S &=& \beta w_z(t) [
k_{yz}(\langle z(t) \rangle_S - \langle y(t) \rangle_S)].
\end{eqnarray}
For the simple case $w_{y}(t) = w_z(t) = \Gamma/N$ this relation
can be directly solved. Here we are particularly interested in the
modification of CMR-II. In analogy to CMR-IInew we use the ansatz
$\dot{g}(t) = \beta k_{xy} w_y(t) g(t)^{K(t)}$ (CMR-II corresponds
to $K(t)=1$). The function $K(t)$ can be directly extracted from
the numerical solution of Eqs.\ref{eqneu} and \ref{eqneu2}. Here
we chose $k_{xy} = k_{yz}$ and $w_{y}(t) = w_z(t) = \Gamma/N$).
Furthermore we determined $W(t)$ by formulating an analogous
equation for the x-particle and solving the resulting three rate
equations in the $\langle . \rangle_0$-ensemble (in analogy to the
procedure in Section 4). Here we chose $w_x(t) = \Gamma$. In Fig.9
we plot the solutions for $K(t)$ and $W(t)$ against each other for
$N=64$. One can see that $K(t)$ decays from 2 to 1 where the decay
is mainly in the region where $W(t)$ is close to its final value.
Thus for a broad range of times CMR-IInew with $K \approx 2$ is
the appropriate rate equation. Interestingly, $K=2$ is a typical
value used for the description of experimental conductivity
spectra \cite{Funke02}. For different ratios $k_{xy}/k_{yz}$
slightly different $K(t)$-dependencies result.

We just mention that a further useful modification is the
substitution of the harmonic potential $k(y_i - y_j)^2$ by the
periodic potential $2(1-\cos(\sqrt{k}(y_i - y_j)))$. In this
potential it is possible to include the physical effect that an
ion may escape its local ionic cage. Actually, this Hamiltonian
has been extensively analysed in a very different context
\cite{ref11,ref12}.

Although we have discussed only a simple Hamiltonian, the
arguments, concerning the range of applicability of the
CMR-equations, were quite general. Thus one would expect that also
for different model systems or even for realistic ion conductors
similar arguments might hold. It may be possible that an
appropriate redefinition of the successful ensemble (retaining
CMR-II) and thus of the physical interpretation of the function
$g(t)$ may cope with CMR-I. Such a redefinition can be found in
very recent work \cite{Fun03} although the theoretical
implications of such a redefinition concerning, e.g., the validity
of CMR-II still have to be worked out. In any event, if such a
redefinition is possible one may hope that the very good
predictions of the CMR-approach would remain. This speculation is
backed by the observation that CMR-I and CMR-II together,
expressed via $W(t)$, agree very well with experimental data and
CMR-II has found a strictly theoretical justification in our
model. Furthermore it is conceivable that the disorder term
$f_{do}(t)$ may have specific properties which render CMR-I valid
for some situations with additional disorder.

Due to the relevance of the CMR-approach in the field of solid ion
conductors it is essential to illuminate its applicability from a
strictly theoretical point of view. The present work may be
regarded as a step in this direction and may hopefully serve as an
input for a future development of the CMR-approach, related, e.g.,
to the interpretation of $K=2$ in the modified CMR-II relation.

We gratefully acknowledge important and helpful conversations with
R.D. Banhatti, K. Funke, and B. Roling.

\clearpage

\begin{list}{}{\leftmargin 2cm \labelwidth 1.5cm \labelsep 0.5cm}

\item[\bf Fig. 1] Sketch of the system of (N+1) ions. The parabola
indicates the range of interaction.

\item[\bf Fig. 2]  Sketch of the definition of the S-ensemble as
the black bars after a jump from -1/2 to 1/2 at t=0. The key idea
is that the fraction of particles which effectively jump to the
left leave the S-ensemble.

\item[\bf Fig. 3] Numerical representation of $W(t)$ and
$W_{CMR}(t)$ together with the numerically determined function
$W(t)$.

\item[\bf Fig. 4]
$1 - g(t)$ vs. -$\ln W(t)$ for $N=64$ and $C = 1.23$.

\item[\bf Fig. 5]
-$\ln g(t)$ vs. $\langle r_y^2(t) \rangle$ for $N=64$ and $C =
1.23$.

\item[\bf Fig. 6]
$1 - g(t)$ vs. -$\ln W(t)$ for $N=64$, $C = 1.23$, and $V_{max} =
2.0$. Scaling works best for $a = 2900$.

\item[\bf Fig. 7]
-$\ln g(t)$ vs. $\langle r_y^2(t) \rangle$ for $N=64$, $C = 1.23$,
and $V_{max} = 2.0$.

\item[\bf Fig. 8]
$\langle \Delta x \rangle_S $, $N\langle \Delta y \rangle_S $, and
$f_{do}(t)$  for $N=64$, $C = 0.07$, and $V_{max} = 2.0 $.

\item[\bf Fig. 9]
$K(t)$ vs. $W(t)$ for the 3-particle system with $N=64$ and
$k_{xy} = k_{yz}$.

\end{list}
\newpage

\begin{figure}
\centerline{\epsfxsize=12cm\epsffile{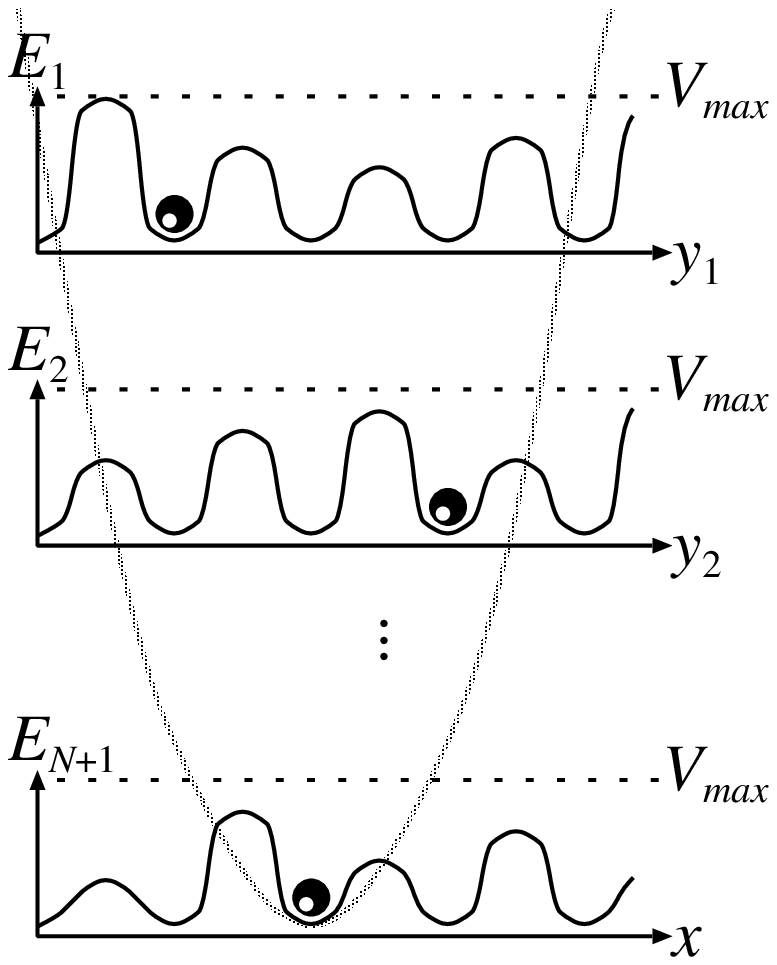}} \centerline{\bf
Fig.\ 1, J. Chem. Phys., M. Kunow et al.}
\end{figure}
\hfill

\newpage

\begin{figure}
\centerline{\epsfxsize=15.5cm\epsffile{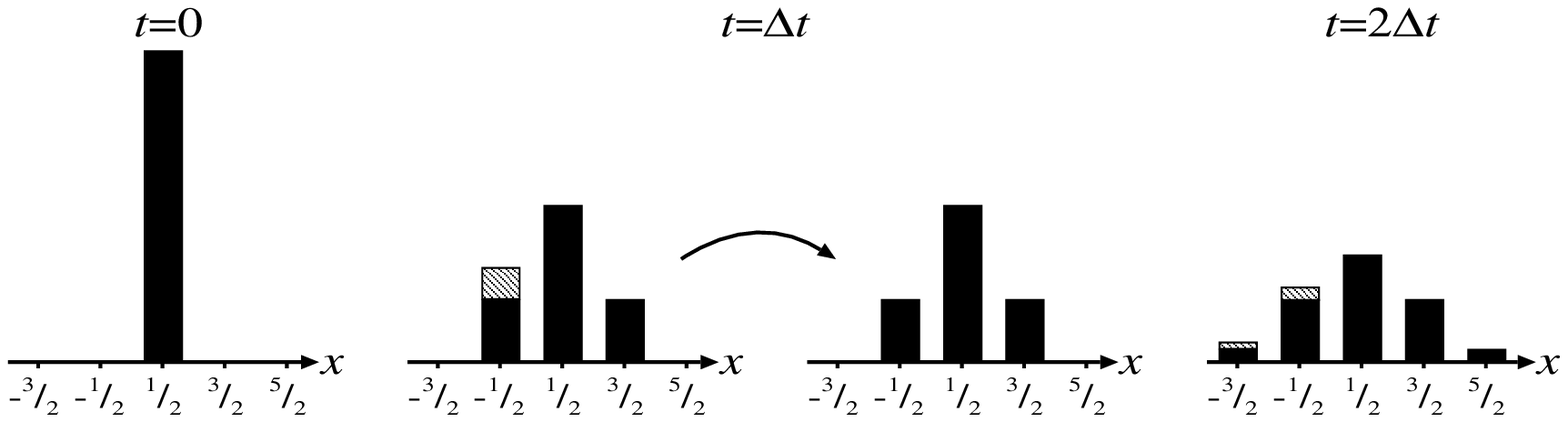}}
\centerline{\bf Fig.\ 2, J. Chem. Phys., M. Kunow et al.}
\end{figure}

\newpage

\begin{figure}
\centerline{\epsfxsize=15.5cm\epsffile{fig_03.eps}}
\centerline{\bf Fig.\ 3, J. Chem. Phys., M. Kunow et al.}
\end{figure}

\newpage

\begin{figure}
\centerline{\epsfxsize=15.5cm\epsffile{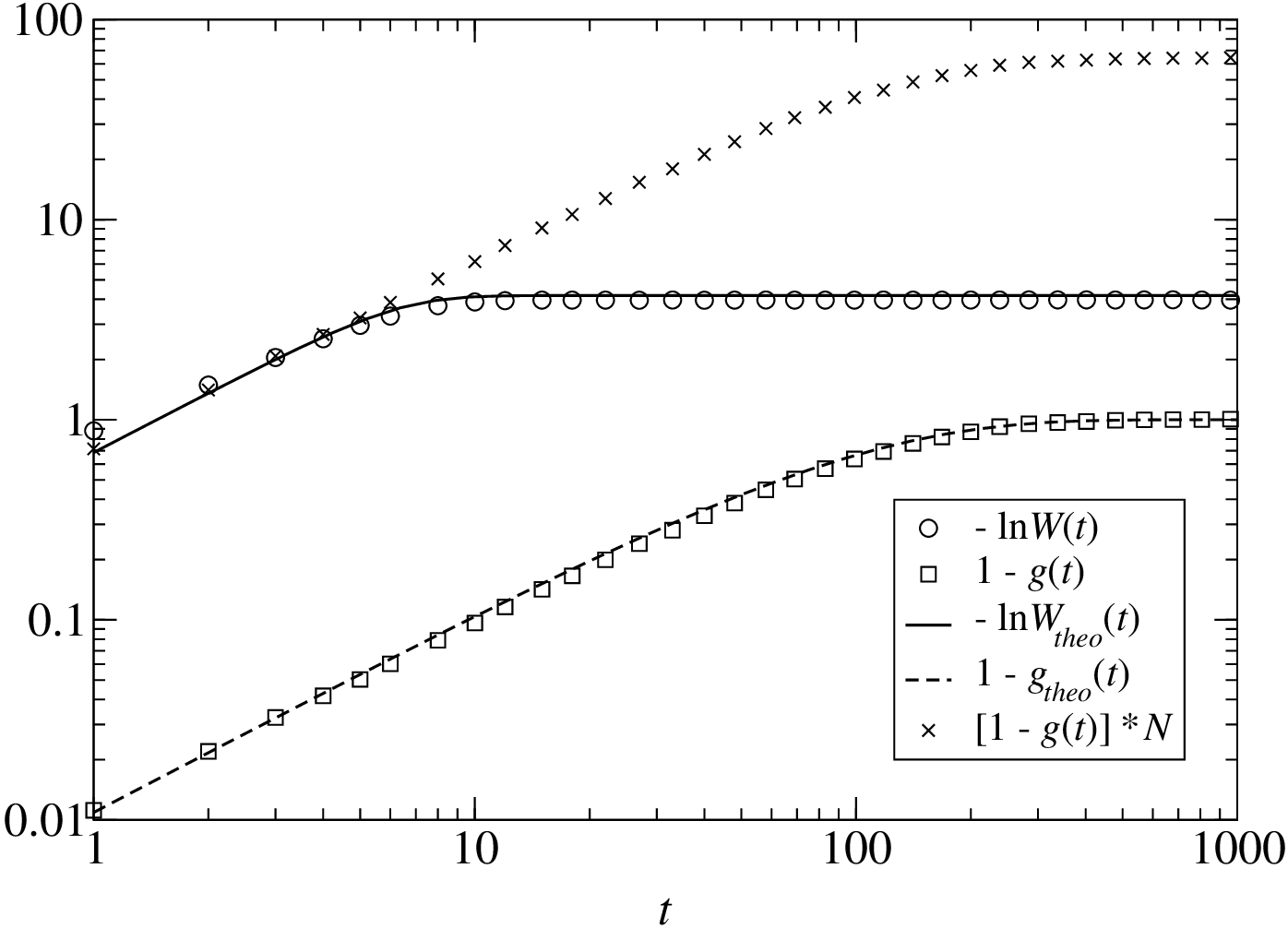}}
\centerline{\bf Fig.\ 4, J. Chem. Phys., M. Kunow et al.}
\end{figure}
\hfill

\newpage

\begin{figure}
\centerline{\epsfxsize=15.5cm\epsffile{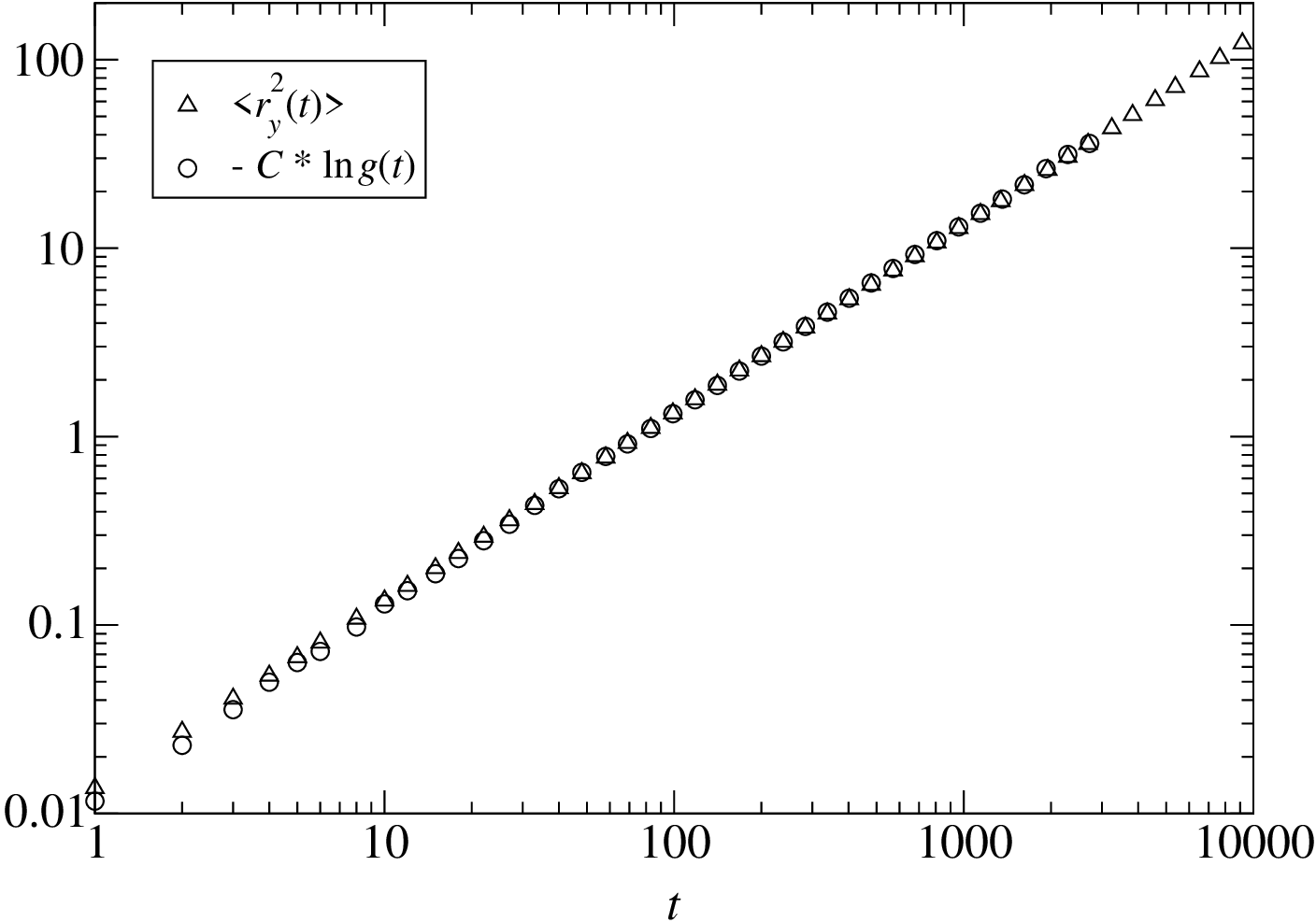}}
\centerline{\bf Fig.\ 5, J. Chem. Phys., M. Kunow et al.}
\end{figure}

\newpage

\begin{figure}
\centerline{\epsfxsize=15.5cm\epsffile{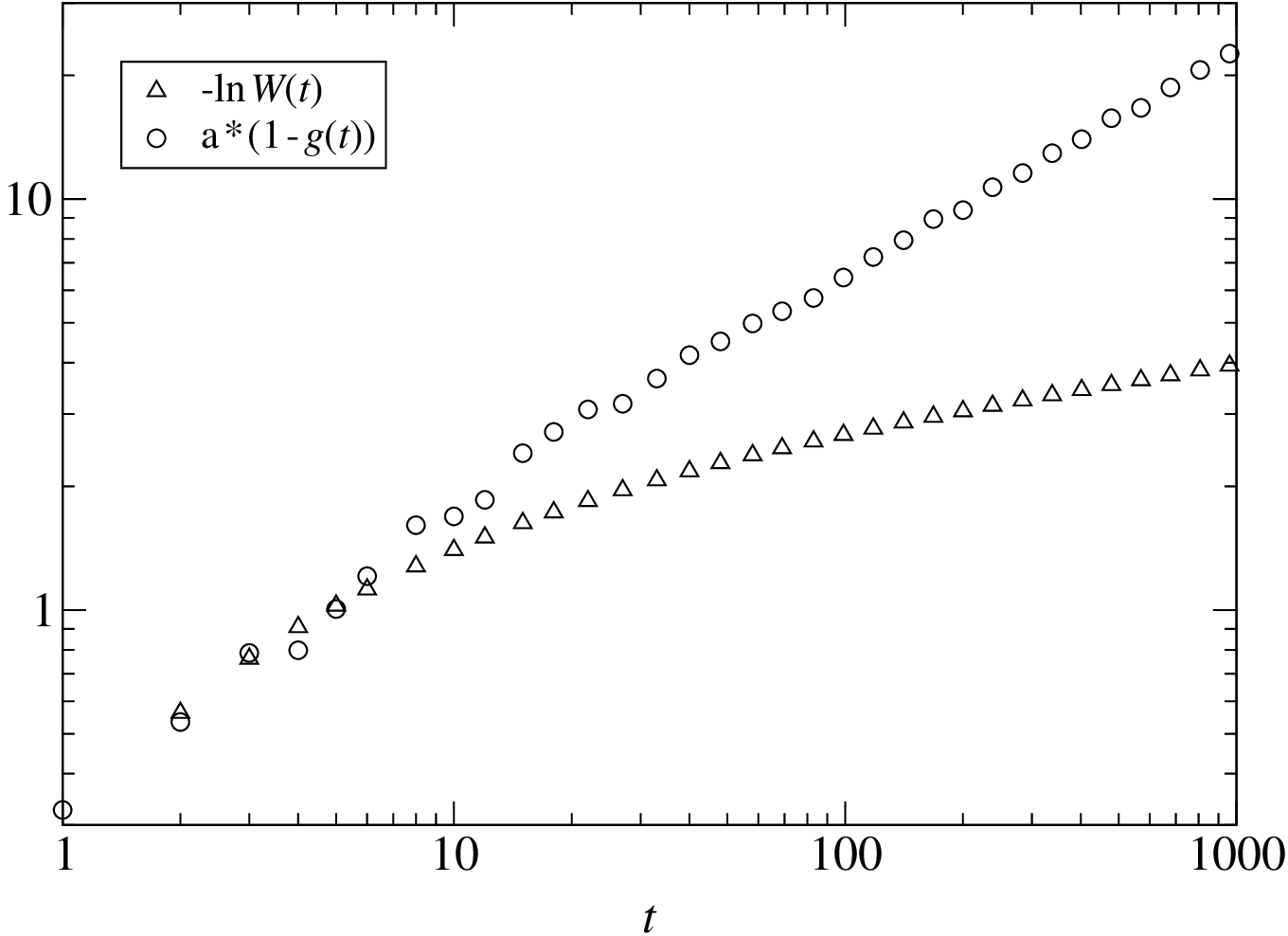}}
\centerline{\bf Fig.\ 6, J. Chem. Phys., M. Kunow et al.}
\end{figure}
\hfill

\newpage

\begin{figure}
\centerline{\epsfxsize=15.5cm\epsffile{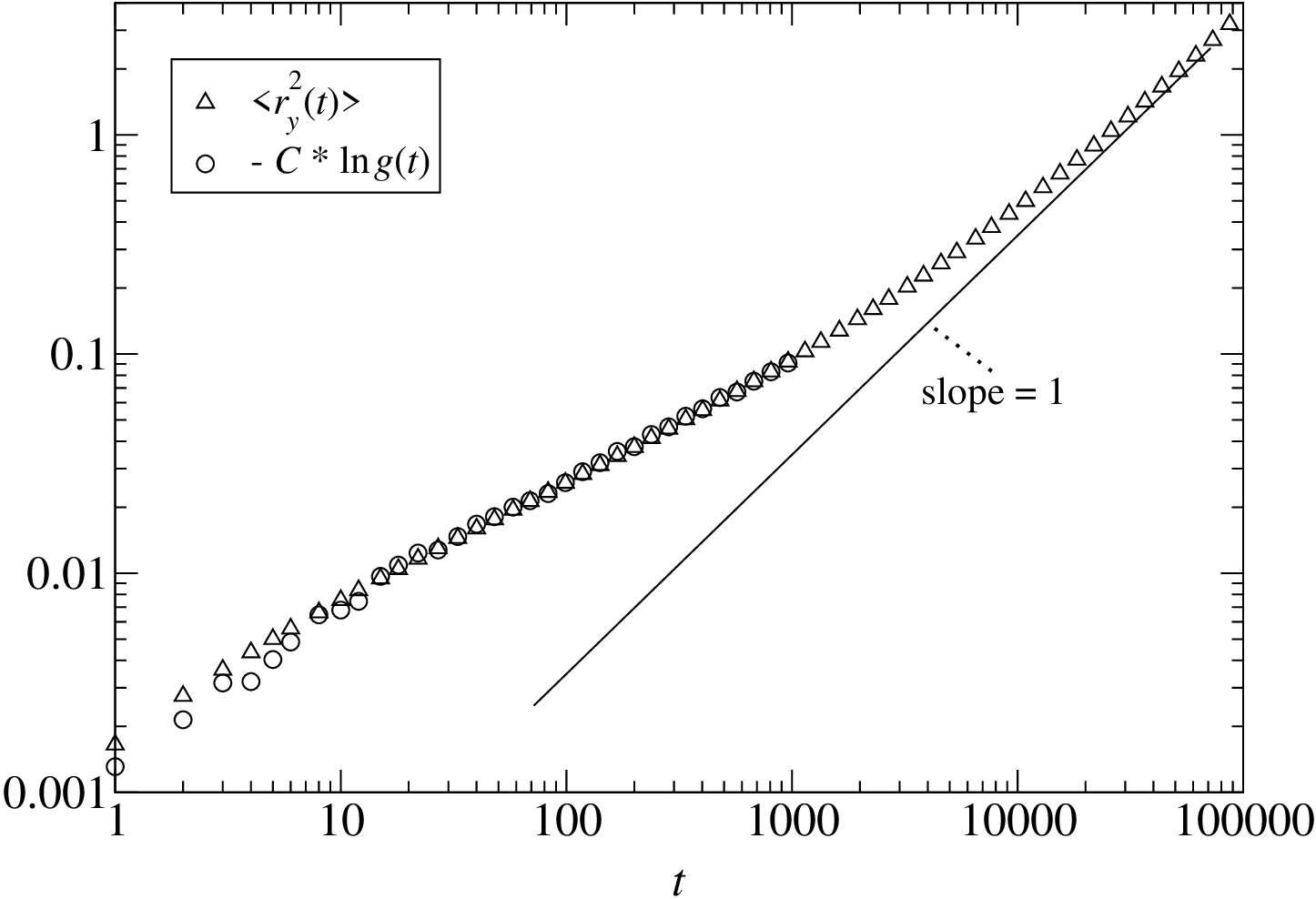}}
\centerline{\bf Fig.\ 7, J. Chem. Phys., M. Kunow et al.}
\end{figure}

\newpage

\begin{figure}
\centerline{\epsfxsize=15.5cm\epsffile{fig_08.eps}}
\centerline{\bf Fig.\ 8, J. Chem. Phys., M. Kunow et al.}
\end{figure}

\newpage

\begin{figure}
\centerline{\rotatebox{270}{\epsfxsize=11.5cm\epsffile{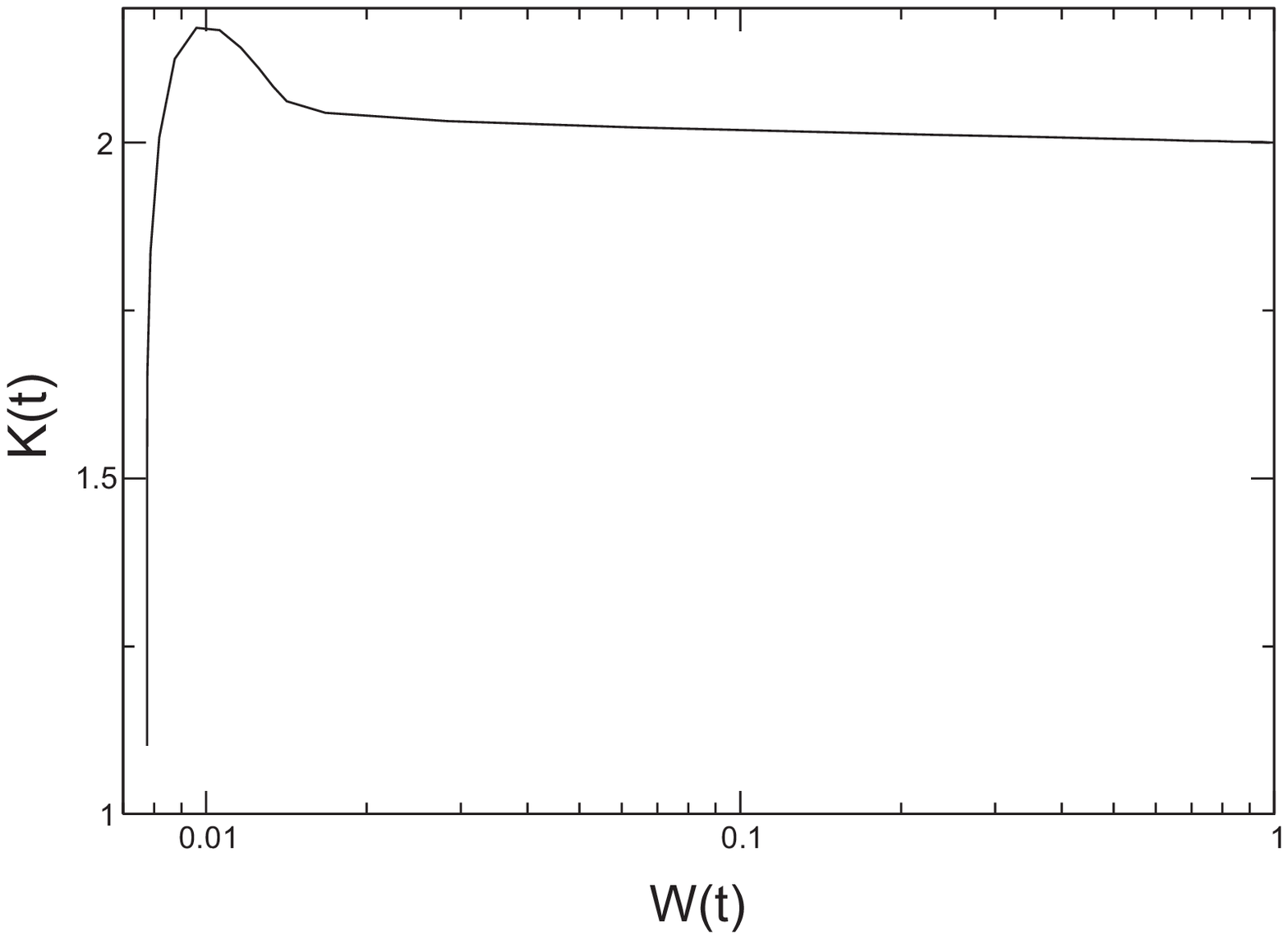}}}
\centerline{\bf Fig.\ 9, J. Chem. Phys., M. Kunow et al.}
\end{figure}

\end{document}